\documentstyle[12pt]{article}
\catcode`@=11
\newcount\@tempcntc
\def\@citex[#1]#2{\if@filesw\immediate\write\@auxout{\string\citation{#2}}\fi
  \@tempcnta\z@\@tempcntb\m@ne\def\@citea{}\@cite{\@for\@citeb:=#2\do
    {\@ifundefined
       {b@\@citeb}{\@citeo\@tempcntb\m@ne\@citea\def\@citea{,}{\bf ?}\@warning
       {Citation `\@citeb' on page \thepage \space undefined}}%
    {\setbox\z@\hbox{\global\@tempcntc0\csname b@\@citeb\endcsname\relax}%
     \ifnum\@tempcntc=\z@ \@citeo\@tempcntb\m@ne
       \@citea\def\@citea{,}\hbox{\csname b@\@citeb\endcsname}%
     \else
      \advance\@tempcntb\@ne
      \ifnum\@tempcntb=\@tempcntc
      \else\advance\@tempcntb\m@ne\@citeo
      \@tempcnta\@tempcntc\@tempcntb\@tempcntc\fi\fi}}\@citeo}{#1}}
\def\@citeo{\ifnum\@tempcnta>\@tempcntb\else\@citea\def\@citea{,}%
  \ifnum\@tempcnta=\@tempcntb\the\@tempcnta\else
   {\advance\@tempcnta\@ne\ifnum\@tempcnta=\@tempcntb \else \def\@citea{--}\fi
    \advance\@tempcnta\m@ne\the\@tempcnta\@citea\the\@tempcntb}\fi\fi}
\catcode`@=12
\newcommand\mini{j_{\rm min}}
\newcommand\one{1\kern-2.5pt{\rm l}}
\newcommand\Frac[2]{\hbox{$\frac{#1}{#2}$}}
\newcommand{\slv}{v\kern-5pt\raise1pt\hbox{$\scriptstyle/$}\kern1pt}
\begin{document}
\begin{flushright}
MZ-TH/96-24 \\[-0.2cm]
August 1996 \\[-0.2cm]
\end{flushright}
\begin{center}

{\Large\bf Constituent Quark Model Approach to Heavy Baryon
Transitions} \\[1.75cm]

{\large J\"urgen G.~K\"orner\footnote{Supported in part by BMBF,FRG
under contract 06MZ566.\\ Invited talk given at the III German-Russian 
Workshop on Heavy Quark Physics, Dubna, Russia, May 20--22, 1996, to appear 
in the Proceedings}} \\[.4cm]
Institut f\"ur Physik, Johannes Gutenberg-Universit\"at \\
Staudingerweg 7, D-55099 Mainz, Germany.

\end{center}
\vspace{1.5cm}
\centerline {\bf ABSTRACT}\noindent
I discuss the structure of current-induced bottom baryon to 
charm baryon transitions, and the structure of pion and photon transitions 
between heavy charm or bottom baryons in the Heavy Quark Symmetry (HQS) limit 
as $m_Q\rightarrow\infty$. By doing a spin-parity analysis I derive a general
formula which allows one to enumerate the independent HQS amplitudes
in the three types of transitions. I go on to show that use of the
constituent quark model for the light-side diquark transitions leads to
a considerable reduction in the number of independent amplitudes derived
in the HQS limit. The discussion includes the ground state $s$-wave
as well as the $p$-wave heavy baryons.

\newpage

\section{Introduction}

In heavy hadrons made up of a heavy quark and a light quark system the 
dynamics of the heavy side and the light side completely decouple in the
heavy mass limit as ${m_Q\rightarrow\infty}$. Since the dynamics of
the heavy quark is known the heavy quark can be viewed as providing a
probe of the unknown light-side dynamics. For example, in current
induced bottom to charm hadron transitions there is a $b\rightarrow c$
transition on the heavy side. The dynamics of this transition is 
completely specified. The light side knows nothing about the heavy-side
transition except for a light-side velocity change.
The velocity change is necessary since the light side quark system has to
readjust its velocity to keep up with the emerging energetic $c-$quark
in order to form the final hadron. There are also some trivial angular
momentum factors that provide
for the requisite projections on the initial and final heavy hadrons
with their given spins and parities. In the heavy flavour conserving pion
and photon transitions the pions and photons are emitted from the light side
and the heavy side knows nothing about these light-side transitions.
It is then an angular momentum coupling exercise to determine the number 
of independent amplitudes that describe the heavy hadron
transitions in the HQS limit. As the dynamics of the heavy-side
transitions is entirely known, all that is needed is to determine
the structure of the light-side transitions.

Turning to heavy baryons made up of a heavy quark and a light diquark 
system one thus has to study the dynamics of the transitions between light
diquark systems very much like one has been studying light ``triquark''
transitions in ordinary light baryon transitions. Many different
approaches of varying degrees of sophistication have been developed
for this purpose in the past. Among these is the constituent quark
model approach which has been singularily succesful in the light baryon
sector. Although one is far from being able to derive the
constituent quark model from first principles one is now beginning to
understand that the constituent quark model may in fact emerge in
the large $N_C$ limit~\cite{dubna1}, at least in the baryon sector.
It is then very tempting and quite natural to try and get a first
handle on the light diquark transitions in the  heavy baryon sector by
using the constituent quark model approach for these. The constituent
quark model predictions will provide simple bench mark predictions for
more sophisticated approaches such as QCD sum rule and lattice 
calculations to be carried out at a later stage.

This defines the aim and the scope of this presentation. In Sec.2 we
provide a brief synopsis of the heavy baryon $s$- and $p$-wave states
expected to arise in a constituent quark model classification. In Sec.3
we provide a general formula which allows one to count the number of
independent reduced amplitudes in current, pion and photon
transitions between heavy baryons in the HQS limit. In Sec.4 we describe
how the constituent quark model leads to a reduction in the number of the 
reduced amplitudes in heavy baryon transitions. We discuss
some explicit examples and give results on the coupling complexity that 
remains after invoking the constituent quark model for the light-side
diquark transitions. We limit our
attention to transitions involving $s$- and $p$-wave states.

\section{Heavy Baryon $s$- and $p$-Wave States}

A heavy baryon is made 
up of a light diquark system $(qq)$ and a heavy quark $Q$. The light 
diquark system has bosonic quantum numbers $j^P$ with total angular 
momentum $j=0,1,2 \dots$ and parity $P=\pm 1$. To each diquark system with 
spin-parity $j^P$ there is a degenerate heavy baryon doublet with 
$J^P=(j\pm\Frac12)^P$ ($j=0$ is an exception). It is important to realize 
that the HQS structure of the heavy baryon states is entirely determined 
by the spin-parity $j^P$ of the light diquark system. The requisite angular 
momentum coupling factors can be read off from the coupling scheme
$J^P\otimes\Frac12^+\Rightarrow J^P$.
Apart from the angular momentum coupling factors the dynamics
of the light system is completely decoupled from the heavy quark.

Next I turn my attention to the question of which low-lying heavy
baryon states can be expected to exist. From our experience with light 
baryons and light mesons we know that one can get a reasonable description 
of the light particle spectrum in the constituent quark model picture. This 
is particularly true for the enumeration of states, their spins and their 
parities. As much as we know up to now, gluon degrees of freedom do not 
seem to contribute to the particle spectrum. It is thus quite natural to 
try the same constituent approach to enumerate the light diquark states, 
their spins and their parities. 

From the spin degrees of freedom of the two light quarks one obtains a 
spin~0 and a spin~1 state. The total orbital state of the diquark system 
is characterized by two angular degrees of freedom which I take to be the 
two independent relative momenta $k=\frac12(p_1-p_2)$ and 
$K=\frac12(p_1+p_2-2p_3)$ that can be formed from the two light 
quark momenta $p_1$ and $p_2$ and the heavy quark momentum $p_3$. 
The $k$-orbital momentum describes relative orbital excitations of the two 
quarks, and the $K$-orbital momentum describes orbital excitations of the 
center of mass of the two light quarks relative to the heavy quark.
The $(k,K)$-basis is quite convenient in as much as it allows one to
classify the diquark states in terms of $SU(2N_f)\otimes O(3)$ 
representations\cite{dubna2}.
In this presentation I limit my attention to the ground
state $s$-wave and excited $p$-wave heavy baryon states as they occur
in the constituent approach to the light diquark excitations. We 
specify to $N_f=2$ for two ($u,d$)-flavours. The $s$-wave states 
$\Lambda_Q(1/2^+)$ and $\{\Sigma_Q(1/2^+,3/2^+)\}$ are in the $10 \otimes 1$ 
representation of $SU(4)\otimes O(3)$. The $K$-type $p$-wave states are in the 
$10 \otimes 3$ representation with the particle content\cite{dubna2,dubna3}
\begin{equation}
\{\Lambda_{QK1}^{**}(1/2^-,3/2^-)\},\ \Sigma_{QK0}^{**}(1/2^-),\
  \{\Sigma_{QK1}^{**}(1/2^-,3/2^-)\ \mbox{and}\
  \{\Sigma_{QK1}^{**}(3/2^-,5/2^-),\nonumber
\end{equation}
The $k$-type $p$-wave states, finally, are in the $6 \otimes 3$ representation
with the particle content\cite{dubna2,dubna3}
\begin{equation}
\{\Sigma_{Qk1}^{**}(1/2^-,3/2^-)\},\ \Lambda_{Qk0}^{**}(1/2^-),\
  \{\Lambda_{Qk1}^{**}(1/2^-,3/2^-)\ \mbox{and}\
  \{\Lambda_{Qk1}^{**}(3/2^-,5/2^-).\nonumber
\end{equation}
HQS doublets are written in curly brackets. Apart from the ground state $s$-wave 
baryons one thus has altogether seven $\Lambda$-type $p$-wave states and 
seven $\Sigma$-type $p$-wave states. This analysis can easily be extended 
to the case $SU(6)\otimes O(3)$ bringing in the strangeness quark in addition.

\section{Generic Picture of Current, Pion and Photon Transitions}

In $b\rightarrow c$ current transitions, and $c\rightarrow c$ pion and photon 
transitions between heavy baryons the heavy-side and light-side transitions 
occur completely independent of each other (they ``factorize'') in the HQS 
limit except for the requirement that the heavy side and the light side have 
the same velocity in the initial and final state, respectively, which are also 
the velocities of the initial and final heavy baryons. The three types
of transitions are depicted in Fig.1. The $b\rightarrow c$ 
current transition induced by the flavour-spinor matrix~$\Gamma$ is hard 
and, accordingly, in general there is a change of velocities
$v_1\rightarrow v_2$, 
whereas there is no velocity change in the pion and photon transitions. 
The heavy-side transitions are completely specified whereas the light-side 
transitions $j_1^{P_1}\rightarrow j_2^{P_2}$,
$j_1^{P_1}\rightarrow j_2^{P_2}+\pi$ and
$j_1^{P_1}\rightarrow j_2^{P_2}+\gamma$ are described by a number of 
amplitudes which parametrize the light-side transitions.
The pion and the photon couple only to the light side. In the case of the
pion this is due to its flavour content. In the case of the
photon the coupling of the photon to the heavy side involves a spin flip
which is doen by $1/m_Q$ and thus the photon couples only to the light
side in the HQS limit. 

The number $N$ of independent reduced HQS amplitudes
can be obtained by performing a light-side helicity or $LS$-amplitude
analysis~\cite{dubna3}. The result is\\[1cm]
{\it current transitions:}\vspace{-.5cm}
\begin{eqnarray}\label{eqn1}
n_1\cdot n_2&=&1\qquad N=\mini+1\nonumber\\
n_1\cdot n_2&=&-1\quad N=\mini
\end{eqnarray}
{\it pion transitions:}\vspace{-.5cm}
\begin{eqnarray}\label{eqn4}
n_1\cdot n_2&=&1\qquad N=\mini\nonumber\\
n_1\cdot n_2&=&-1\quad N=\mini+1
\end{eqnarray}
{\it photon transitions:}\vspace{-.5cm}
\begin{eqnarray}\label{eqn5}
j_1&=&j_2\qquad N=2j_1\nonumber\\
j_1&\neq&j_2\quad N=2\mini+1
\end{eqnarray}
In the case of current and pion transitions the
counting involves the normalities of the light-side diquarks which are
defined by $n=(-1)^jP$.

\section{Constituent Quark Model Approach to Light-Side Transitions}

Interest in the constituent quark model has recently been rekindled by the 
discovery (or rediscovery) that two-body spin-spin interactions between 
quarks are non-leading in $1/N_C$, at least in the baryon 
sector~\cite{dubna1}. Thus, to leading order in $1/N_C$, light quarks 
behave as if they were heavy as concerns their spin interactions.
In the constituent quark model approach one further assumes that spin and 
orbital degrees of freedom decouple. One can therefore classify the light
diquark system in terms of $SU(2N_f) \otimes O(3)$ symmetry multiplets.
Transitions between light quark systems are parametrized in terms of
a set of one-body operators whose matrix elements are then evaluated 
between the $SU(2N_f) \otimes O(3)$ multiplets~\cite{dubna2,dubna4,dubna5}.

Let us illustrate this for the $(b\rightarrow c)$ current induced ground
state to ground state transitions. For these there are altogether three 
reduced HQS form factors or Isgur-Wise functions, one for the 
$\Lambda_b\rightarrow\Lambda_c$ transition and two for the 
$\{\Sigma_b\}\rightarrow\{\Sigma_c\}$ transitions. The ground state to
ground state transition is simple in that there is only one one-body
operator. Thus the three reduced HQS form factors can all be expressed 
in terms of a single form factor $A(\omega)$, where $A(1)=1$ at zero 
recoil. One then finds that the current transition amplitudes are given 
by~\cite{dubna2,dubna3,dubna6}
\begin{eqnarray}
\Lambda_b\rightarrow\Lambda_c
  &:&M^\lambda=\bar u_2\Gamma^\lambda u_1\frac{\omega+1}2A(\omega)\\
  \{\Sigma_b\}\rightarrow\{\Sigma_c\}
  &:&M^\lambda=\bar\psi_2^\nu\Gamma^\lambda\psi_1^\mu
  (-\frac{\omega+1}2g_{\mu\nu}+\frac12v_1^\nu v_2^\mu)A(\omega)\nonumber
\end{eqnarray}
The same result has been obtained by C.K.Chow by analyzing the large
$N_C$ limit of QCD~\cite{dubna7}.

For the current transitions from the bottom baryon ground states to the
$p$-wave charm baryon states one similarly reduces the number of reduced 
form factors when using the constituent quark model in addition to HQS. 
For the transition into the $K$-multiplet one has a reduction from five 
HQS reduced form factors to two constituent quark model form factors  
whereas for transitions into the $k$-multiplet one can relate two HQS 
reduced form factors to one single spin-orbit form
factor~\cite{dubna2}. These are
testable predictions in as much as the population of helicity states in
the daughter baryon is fixed resulting in a characteristic decay pattern
of its subsequent decay.

The one-pion and photon transitions can be 
treated in a similar manner. Again one finds a significant simplification 
of the HQS structure, i.e. the number of independent HQS amplitudes
is reduced from 
those listed in Eqs.~(\ref{eqn4}) and~(\ref{eqn5}) when the constituent 
quark model is invoked in addition to HQS. Results for the
one-pion transitions can be found in~\cite{dubna4} and
for the photon transitions in ~\cite{dubna5}. We summarize our results
in Table1 where we list the number of independent reduced HQS
amplitudes together with the number of independent amplitudes that
remain when the constituent quark model is invoked. 
\begin{table}
\begin{center}
\begin{tabular}{|l|c|c|}
\hline
&&constituent\\
&HQS&quark model\\
\hline
{\it Current transitions:}&&\\
$s$-wave to $s$-wave&3&1\\
$s$-wave to $p$-wave ($K$)&5&2\\
$s$-wave to $p$-wave ($k$)&2&1\\
\hline
{\it Pion transitions:}&&\\
$s$-wave to $s$-wave&2&1\\
$p$-wave ($K$) to $s$-wave&7&2\\
$p$-wave ($k$) to $s$-wave&5&2\\
\hline
{\it Photon transitions:}&&\\
$s$-wave to $s$-wave&3&1\\
$p$-wave ($K$) to $s$-wave&11&2\\
$p$-wave ($k$) to $s$-wave&11&2\\
\hline
\end{tabular}
\end{center}
\caption{Number of independent amplitudes for current, pion and photon 
transitions between heavy baryons in the HQS limit and in the constituent 
quark model\label{tab1}}
\end{table}
We mention that the dramatic reduction in the number of independent
amplitudes shown in Table1 is specific to the heavy baryon sector and
is no longer true for heavy mesons.

\vspace{1cm}
\noindent{\bf Acknowledgement:}\\
Much of the material presented in this review is drawn from work done in 
collaboration with F.~Hussain, M.~Kr\"amer, J.~Landgraf, D.~Pirjol and 
S.~Tawfiq. I would like to thank them for their collaborative efforts.
My thanks are also going to S.~Groote
for help and advice on the LATEX version of this report.

\vspace{1cm}
\centerline{\Large\bf Figure Captions}
\vspace{.5cm}
\newcounter{fig}
\begin{list}{\bf\rm Fig.\ \arabic{fig}:}{\usecounter{fig}
\labelwidth1.6cm\leftmargin2.5cm\labelsep.4cm\itemsep0ex plus.2ex}

\item Generic picture of bottom to charm current transitions, and 
  pion and photon transitions in the charm sector in the HQS limit 
  $m_Q\rightarrow\infty$

\end{list}

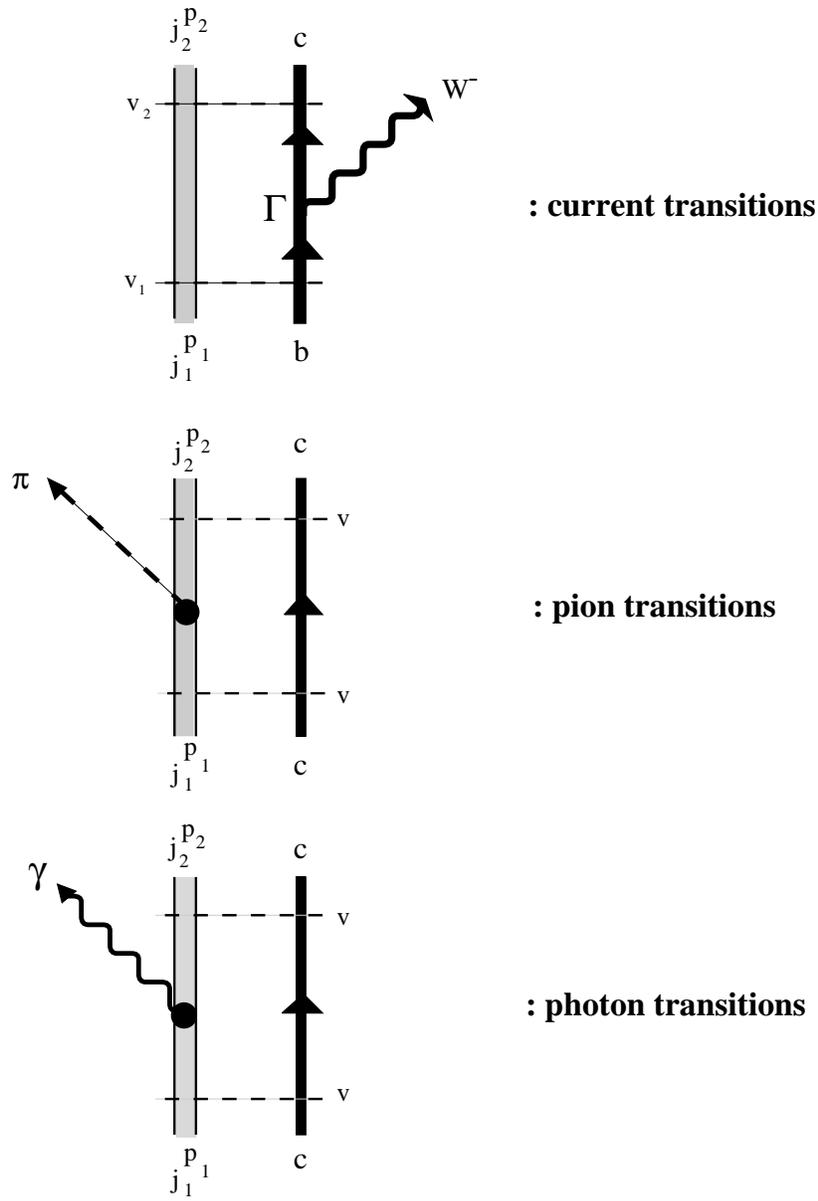
\end{document}